\pdfoutput=1

\documentclass[11pt]{article}

\usepackage{acl}

\usepackage{times}
\usepackage{latexsym}

\usepackage[T1]{fontenc}

\usepackage[utf8]{inputenc}

\usepackage{microtype}

\usepackage{multirow}
\usepackage{tikz}
\usepackage{pgfplots}
\usepackage{pgfplotstable}
\usepackage{subcaption}
\usepackage{tabularx}
\usepackage{layouts}
\usepackage{amsfonts}

\title{Learning Diverse Document Representations with Deep Query Interactions for Dense Retrieval}

\author{Zehan Li\thanks{\ \ Work done during internship at Microsoft Research Asia.} \\
  Beihang University \\
  \texttt{lizehan@buaa.edu.cn} \And
  Nan Yang \\
  Microsoft Research Asia \\
  \texttt{nanya@microsoft.com} \AND
  Liang Wang \\
  Microsoft Research Asia \\
  \texttt{wangliang@microsoft.com} \And
  Furu Wei \\
  Microsoft Research Asia \\
  \texttt{fuwei@microsoft.com}}

\begin{document}
\maketitle
\begin{abstract}

In this paper, we propose a new dense retrieval model which learns diverse document representations with deep query interactions.
Our model encodes each document with a set of generated pseudo-queries to get query-informed, multi-view document representations.
It not only enjoys high inference efficiency like the vanilla dual-encoder models, but also enables deep query-document interactions in document encoding and provides multi-faceted representations to better match different queries.
Experiments on several benchmarks demonstrate the effectiveness of the proposed method, out-performing strong dual encoder baselines.\footnote{The code is available at \url{https://github.com/jordane95/dual-cross-encoder}.}

\end{abstract}

\section{Introduction}

Document retrieval plays an important role in information retrieval (IR) tasks such as web search and open domain question answering \citep{chen-etal-2017-reading}.
Early works such as BM25-based retriever \citep{DBLP:journals/ftir/RobertsonZ09} rely on lexical term matching to calculate the relevance of a pair of texts.
Recently, neural network based dense retrieval \citep{karpukhin-etal-2020-dense} has gained traction in research community. Dense retrieval learns a neural encoder to map queries and documents into a dense, low-dimensional vector space, and is less vulnerable to term mismatch problem compared to lexical match-based approaches.

There are two architectures to model the relevance between queries and documents.
Dual encoder architecture encodes query and document separately into fixed-dimensional vectors \citep{karpukhin-etal-2020-dense}, where
the similarity between query and document is usually instantiated as a dot product or cosine similarity between their vectors. As there are no interactions between query and document, dual encoder approach permits efficient inference with vector space search on pre-computed document vectors.
Cross encoder architecture feeds the concatenation of a query and document pair into one encoder to calculate its relevance score \citep{DBLP:journals/corr/abs-1901-04085}.
Compared to dual encoder, cross encoder is more accurate due to the deep interaction between query and document, but comes with computation costs infeasible for first-stage retrieval. It is highly desirable to design a retrieval model which can match the performance of the cross encoder approach while maintaining the inference efficiency of the dual encoder approach. 

To this end, previous works mainly focus on two directions: late-interaction and distillation.
The first solution is to design a hybrid architecture, where the early layers act as a dual encoder while the late layers work like a cross encoder \citep{DBLP:conf/sigir/MacAvaneyN0TGF20,  DBLP:conf/sigir/KhattabZ20, Humeau2020Poly-encoders:}.
Its effectiveness comes with the cost of retrieval latency due to the computation involved with late layers.
Another solution is knowledge distillation \citep{DBLP:journals/corr/HintonVD15}, using the cross encoder to augment the training data \citep{qu-etal-2021-rocketqa, ren-etal-2021-rocketqav2},
or distilling the ranking scores or attention matrices of a more powerful cross encoder reranker to a dual encoder retriever \citep{DBLP:conf/sigir/HofstatterLYLH21, ren-etal-2021-rocketqav2, DBLP:journals/corr/abs-2205-09153}.

\begin{figure*}[!ht]
\centering
\tikzset{
de/.style ={
    rectangle,
    rounded corners = 5pt,
    minimum height = 40pt,
    inner sep=5pt,
    draw=gray,
    fill = lightgray!10,
    align = center,
},
ce/.style = {
    rectangle,
    rounded corners =5pt,
    minimum width = 110pt,
    minimum height =40pt,
    inner sep=5pt,
    draw = gray,
    fill = gray!20,
    align = center,
},
query/.style = {
    rectangle,
    rounded corners = 3pt,
    minimum height = 20pt,
    draw = gray,
    fill = pink!15,
    align = center,
},
doc/.style = {
    rectangle,
    rounded corners = 3pt,
    minimum height = 20pt,
    draw = gray,
    fill = red!15,
    align = center,
},
cls/.style = {
    rectangle,
    minimum width = 3pt,
    minimum height = 1pt,
    draw = gray,
    fill = yellow!50,
    align = center,
    font=\small,
},
}
\begin{subfigure}{0.25\textwidth}
    \centering
    \begin{tikzpicture}
    \node[query] (q) at (0, 0.5) {Query};
    \node[doc] (d) at (2, 0.5) {Document};
    \node[de] (qe) at (0,2) {Query \\ Encoder};
    \node[de] (de) at (2,2) {Document \\ Encoder};
    \node[cls] (qcls) at (0, 3.5) {CLS};
    \node[cls] (dcls) at (2, 3.5) {CLS};
    \node (score) at (1, 4.5) {score};
    \draw[-latex] (q) -- (qe);
    \draw[-latex] (d) -- (de);
    \draw[-latex] (qe) -- (qcls);
    \draw[-latex] (de) -- (dcls);
    \draw[-latex] (qcls) -- (score);
    \draw[-latex] (dcls) -- (score);
    \end{tikzpicture}
    \caption{Dual Encoder}
    \label{fig:dual-encoder}
\end{subfigure}
\hfill
\begin{subfigure}{0.25\textwidth}
    \centering
    \begin{tikzpicture}
    \node[query] (q) at (0, 0.5) {Query};
    \node[doc] (d) at (2, 0.5) {Document};
    \node[ce] (ce) at (1, 2) {Cross Encoder};
    \node[de, minimum width = 5pt, minimum height = 5pt] (mlp) at (1, 3.5) {MLP};
    \node (score) at (1, 4.5) {score};
    \draw[-latex] (q.north) -- (q.north|-ce.south);
    \draw[-latex] (d.north) -- (d.north|-ce.south);
    \draw[-latex] (ce) -- (mlp) -- (score);
    \end{tikzpicture}
    \caption{Cross Encoder}
    \label{fig:cross-encoder}
\end{subfigure}
\hfill
\begin{subfigure}{0.4\textwidth}
    \centering
    \begin{tikzpicture}
    \node[query] (q) at (0, 0.5) {Query};
    \node[query] (qq) at (2, 0.5) {Query};
    \node[doc] (d) at (4, 0.5) {Document};
    \node[de] (qe) at (0, 2) {Query \\ Encoder};
    \node[ce] (ce) at (3, 2) {Document \\ Encoder};
    \node[cls] (qcls) at (0, 3.5) {CLS};
    \node[cls] (dcls) at (3, 3.5) {CLS};
    \node (score) at (1.5, 4.5) {score};
    \draw[-latex] (q) -- (qe);
    \draw[-latex] (qq.north) -- (qq.north|-ce.south);
    \draw[-latex] (d.north) -- (d.north|-ce.south);
    \draw[-latex] (qe) -- (qcls);
    \draw[-latex] (ce) -- (dcls);
    \draw[-latex] (qcls) -- (score);
    \draw[-latex] (dcls) -- (score);
    \end{tikzpicture}
    \caption{Dual Cross Encoder}
    \label{fig:dual-cross-encoder}
\end{subfigure}

\caption{Illustration of different matching paradigms with different architectures.}
\label{fig:encoders}
\end{figure*}
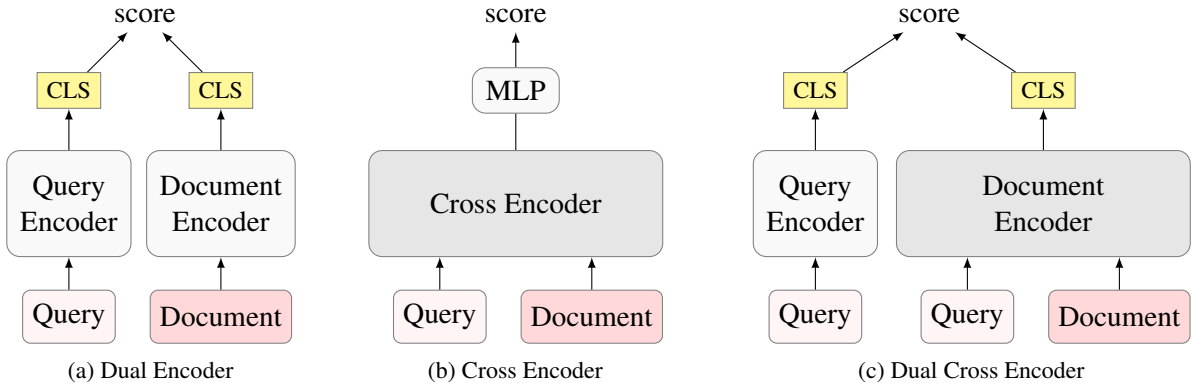

In this paper, we propose to achieve this goal by pre-computing the interaction-based representations.
As depicted in Figure~\ref{fig:dual-cross-encoder}, the document representations are obtained by feeding the concatenation of query and document through a cross encoder while the query representation is obtained in the same way as in the vanilla dual encoder.
For every document, we use a query generation model to generate several queries which will each concatenate with the document to obtain a separate document representation. 

Our model has the following advantages.
Firstly, we can obtain document representation with deep query interactions without much additional inference cost.
Additionally, we can naturally get multi-view document representations \citep{luan-etal-2021-sparse, tang-etal-2021-improving-document, zhang-etal-2022-multi} by treating the query as explicit view extractor.

We follow the popular contrastive learning paradigm for learning such representations.
Experiments on several retrieval benchmarks demonstrate the effectiveness of the proposed approach.

To summarize, our main contributions are as follows:
\begin{itemize}
    \item We propose a new model architecture for dense retrieval, which can benefit from deep query-document interaction with low inference latency and learn multi-view document representations to better match different queries.
    \item We show the effectiveness of this model over various baselines by experiments on several large-scale retrieval benchmarks.
\end{itemize}

\section{Related Work}

\subsection{Dense Retrieval}

Dense passage retrieval (DPR) \citep{karpukhin-etal-2020-dense} learns a two-tower BERT encoder to represent question and passage as vectors and takes their dot product as relevance score.
The training of such dense retrievers can be optimized with more sophisticated negative sampling strategy \citep{xiong2021approximate, qu-etal-2021-rocketqa, DBLP:conf/sigir/HofstatterLYLH21, DBLP:conf/sigir/ZhanM0G0M21, yang-etal-2021-xmoco},
or knowledge distillation from a more powerful cross-encoder teacher \citep{qu-etal-2021-rocketqa, ren-etal-2021-rocketqav2, DBLP:conf/sigir/HofstatterLYLH21, DBLP:journals/corr/abs-2205-09153}.

Recently, some work have been devoted to trading off the efficiency and effectiveness with a late-interaction architecture.
\citet{Humeau2020Poly-encoders:} compress the query context into multiple dense vectors with a Poly-Encoder architecture. The relevance score is modeled by a attention-weighted sum of individual matching scores.
\citet{tang-etal-2021-improving-document} further improve the multi-encoding scheme through $k$-means clustering over all document tokens' embeddings.
ColBERT \citep{DBLP:conf/sigir/KhattabZ20} learns word level representations for both query and document and calculates the relevance score with a MaxSim operation followed by a sum pooling aggregator.
Although powerful, they cannot fully utilize the maximum inner product search (MIPS).
In contrast, we employ a pre-interaction mechanism combined with a max pooler which is compatible with MIPS.

Multi-vector encoding is essential in these late-interaction models, but is also gradually borrowed to learn effective dense retrieval models.
\citet{luan-etal-2021-sparse} represent each document with its first $k$ token embeddings.
To learn multi-view document representations, \citet{zhang-etal-2022-multi} substitute the \verb|[CLS]| token with $k$ special \verb|[VIE]| tokens as view extractors and propose a local contrastive loss with annealing temperature between different views.
In comparison, our model learns diverse document representations through interactions with different queries.

\subsection{Query Generation}

Query generation (QG) is originally introduced to the IR community as a document expansion technique \citep{DBLP:journals/corr/abs-1904-08375}.
\citet{docT5query} show that appending the T5-generated queries to the document before building the inverted index can bring substantial improvements over BM25.
More recently, \citet{DBLP:conf/sigir/MalliaKST21} use generated queries as term expansion to learn better sparse representations for documents.

In the context of dense retrieval, query generation is usually used for domain adaptation in data scarcity scenarios.
For example, \citet{ma2020zero} use QG model trained on general domain to generate synthetic queries on target domain for model training.
To reduce noise in generated data, \citet{wang-etal-2022-gpl} further introduce a cross encoder for pseudo labeling.
Different from the previous work, we mainly use the generated queries to learn query-informed document representations.

\section{Method}

\subsection{Task Definition}

Given a query $q$ and a collection of $N$ documents $\mathcal{D}=\{d_1,d_2,...,d_i,...,d_N\}$, a retriever aims to find a set of $K$ relevant documents $\mathcal{D}_+=\{d_{i_1},d_{i_2},...,d_{i_j},...,d_{i_K}\}$,\footnote{Usually $K\ll N$.}
by ranking the document in the corpus $\mathcal{D}$ according to its relevance score with respect to the query $q$, for next stage re-ranking or downstream applications.

\subsection{Dual Encoder}

We first introduce the dual encoder (DE) architecture for dense retrieval.
In this framework, a query encoder $DE_q$ and a document encoder $DE_d$ are used to encode the query and document into low-dimensional vectors, respectively.
To measure their relevance, a lightweight dot product between the two vectors is usually adopted to enable fast search,
\begin{equation}
    s(q,d)=DE_q(q)\cdot DE_d(d).
\end{equation}


The common design choice for the encoders is using multi-layer Transformers \citep{NIPS2017_3f5ee243} initialized from pre-trained language models (PLMs), such as BERT \citep{devlin-etal-2019-bert}.
How to get the representation from BERT is also an interesting question but beyond the scope of this paper.
For simplicity, we directly take the \verb|[CLS]| vector at the final layer as the text representation.
The two encoders can share or use separate parameters.
We tie the encoder parameters in main experiments but also provide results of untied parameters in ablation study.

\subsection{Cross Encoder}

The cross encoder (CE) takes the concatenation of query and document as input and uses deep neural network to model their deep interactions.
Given a pair of query and document consisting of multiple tokens, we feed their concatenation through a cross encoder to get the interaction-aware representation,
\begin{equation}
    \mathbf{r}=CE(q+d).
\end{equation}
Then a multi-layer perceptron (MLP) is applied on top of the interaction-aware representation to predict the relevance score,
\begin{equation}
    s(q,d) = MLP(\mathbf{r}).
\end{equation}

The cross encoder is also usually instanced as a multi-layer Transformer network initialized from BERT.
It can model term-level interactions between query and document, providing more fine-grained relevance estimation.

\subsection{Dual Cross Encoder}

We present our dual cross encoder where the document encoder acts as a cross encoder whereas the query encoder works like a dual encoder.
Specifically, the query representation and document representation with query interaction are calculated as
\begin{equation}\label{eq:qe}
    \textbf{q}=DE_q(q),
\end{equation}
\begin{equation}\label{eq:qd}
    \textbf{d}=CE_d(q+d).
\end{equation}
Their similarity is measured by a dot product like in the vanilla dual encoder,
\begin{equation}
    s(q,d)=\textbf{q}\cdot \textbf{d}.
\end{equation}

\paragraph{Query Generation.}
Note that the query from the query encoder side and document encoder side do not necessarily have to be the same since we only have access to the gold query for documents appearing in the training set.
It is impractical to manually write potential queries for each document in the whole corpus.
Hence, we use a T5 model \citep{JMLR:v21:20-074} fine-tuned on the doc-to-query task to generate queries for each document.
We empirically adopt 10 queries decoded with top-$k$ sampling strategy \citep{fan-etal-2018-hierarchical} to encourage the query generation diversity.

The advantages of this architecture are two-fold.
On the one hand, we can model the query-document interaction in the pre-computed document representations.
On the other hand, we can enjoy the retrieval efficiency of the vanilla dual encoder by pre-computing the interaction-aware document representations.

\subsection{Training}
\label{sec:training}
The conventional way to train a dense retriever requires a set of $(q,d_+,d_-)$ pairs.
The model is trained by optimizing the contrastive loss,
\begin{equation}
    L(q,d_+,\mathcal{D}_-) = - \log\frac{e^{s(q,d_+)}}{\sum\limits_{d\in \{d_+\}\cup\mathcal{D}_-}e^{s(q,d)}},
\end{equation}
where $ \mathcal{D}_-$ contains a set of negative documents $d_-$ for query $q$.
Following \citet{karpukhin-etal-2020-dense}, we include both BM25 hard negatives and in-batch negatives in $ \mathcal{D}_-$.

\paragraph{Constructing Positives and Negatives.}
Fusing query information into document representation requires redefining the positive and negative pairs.
For a given query $q$, our framework potentially permits four types of positive and negatives, namely, $(q_++d_+),(q_++d_-),(q_-+d_+)$ and $(q_-+d_-)$.
To train our model, we convert the traditional positive and negative pair from the training set into that in our framework with the mapping function
\begin{equation}
    f: (q,d_+,d_-) \mapsto (q,q+d_+,q+d_-),
\end{equation}
where the $+$ is the concatenation operation used in cross encoder.
This mapping leads to the positive of type $(q_++d_+)$ and the following two types of negatives.
We leave the exploration of other types of negatives to future work.

\paragraph{Hard Negatives.}
The negative documents $d_-$ are usually randomly sampled from BM25 top-ranked documents.
After the mapping function defined above, these negatives fall in the type of negative $(q_++d_-)$, which serve as hard negatives in our framework.
This type of negatives can teach the model to learn more fine-grained information, as $d_-$ is usually topically related to the gold query but cannot exactly answer the query.
It also prevents our model from learning the shortcut, \emph{i.e.}, only learning matching signals from the query side and ignoring the document side information.

\paragraph{In-Batch Negatives.}
To improve the training efficiency, we also adopt in-batch negatives to train our model.
In our framework, the in-batch negatives belong to the negative type $(q_-+d_-)$.
This type of negatives is simple and can enable the model to learn topic-level discrimination ability.

\paragraph{Data Augmentation.} 
Regarding the generated queries as weakly annotated data, we can first pre-train our model on these noisy data as a warm-up stage and then fine-tune it on the human-annotated high-quality training set.

\subsection{Inference}

\paragraph{Index} We encode the corpus following the same format as Equation~\ref{eq:qd}, to get multi-view document representations with deep query interactions.

Denoting $\mathbf{d}_j^i$ as the $i$-th view of the $j$-th document $d_j\in\mathcal{D}$, we have
\begin{equation}
    q_j^i\sim P_{QG}(q|d_j),
\end{equation}
\begin{equation}
    \mathbf{d}_j^i = CE_d(q_j^i+d_j),
\end{equation}
where $P_{QG}(q|d)$ denotes the query generation model, $i\in\{1,...,k\}$ and $j\in\{1,...,N\}$.

\paragraph{Retrieval} When a query comes, we encode it with the query encoder to get its contextualized representation $\mathbf{q}$ as in Equation~\ref{eq:qe}.
We adopted multi-vector encodings for a document $d_j$, the relevance score between the query $q$ and the document $d_j$ is taken as the max pooling of its different views' scores,
\begin{equation}
    s(q,d_j) = \max_i \mathbf{q}^T\mathbf{d}_j^i.
\end{equation}
This operation is compatible with MIPS for efficiency optimization,\footnote{Note that to get top-$K$ documents, we first retrieve $10K$ documents to ensure that we have at least $K$ documents after pooling.}
\begin{equation}
    p = \arg \max_d \mathbf{q}^T \mathbf{d}.
\end{equation}

\section{Experiment}

In this section, we evaluate our model on different retrieval benchmarks and compare it with various baselines.

\subsection{Datasets}

We conduct experiments on the following retrieval benchmarks.

\textbf{MS MARCO} is a retrieval benchmark that originates from a machine reading comprehension dataset containing real user queries collected from Bing search and passages from web collection \citep{https://doi.org/10.48550/arxiv.1611.09268}.
We evaluate our model on the passage retrieval task.
The corpus contains about 8.8M passages.
The training set consists of about 500k annotated query-document pairs.
The dev set has 6980 annotated queries.
Since the test set is not publicly available, we evaluate on the dev set following previous work.

\textbf{TREC} Deep Learning (DL) tracks provide test sets with more elaborate annotations to evaluate the real capacity of ranking models.
We evaluate on the 2019 and 2020 test set \citep{DBLP:journals/corr/abs-2003-07820, DBLP:conf/trec/CraswellMMYC20}.
The 2019 test set contains 43 annotated queries and the 2020 test set contains 54 annotated queries.
Both of them share the same corpus with the MS MARCO passage retrieval benchmark.

\subsection{Evaluation Metrics}
Following previous work, we mainly evaluate the retrieval performance on MS MARCO passage retrieval benchmark with MRR@10 but also report the score of Recall@1000.
For datasets from TREC DL tracks, we evaluate with nDCG@10.

\subsection{Baselines}

We mainly compare our model against the DPR \citep{karpukhin-etal-2020-dense} baseline with a dual encoder architecture, but also report results of the following models most related to ours.

\begin{itemize}
    \item BM25 \citep{DBLP:journals/ftir/RobertsonZ09} is the traditional lexical retriever.
    \item DocT5Query \citep{docT5query} appends generated queries to the document before building the inverted index of BM25.
    \item DeepImpact \citep{DBLP:conf/sigir/MalliaKST21} learns sparse representation for documents using generated queries as expanded terms.
    \item ANCE \citep{xiong2021approximate} trains the DPR model with iterative hard negative mining strategy. We include this baseline since this technique is used in ME-BERT and DRPQ.
    \item ME-BERT \citep{luan-etal-2021-sparse} utilizes the first $k$ token embeddings as multi-vector encodings for documents and adopts max pooling for score aggregation.
    \item DRPQ \citep{tang-etal-2021-improving-document} improves over ME-BERT by performing a $k$-means over all tokens' embeddings and utilizing a attention-based score aggregator.
    \item ColBERT \citep{DBLP:conf/sigir/KhattabZ20} represents query and document at token-level and uses a MaxSim pooler followed by a sum aggregator to calculate the relevance score.
\end{itemize}

\subsection{Implementation}

We implement our model based on the \verb|tevatron| toolkit \citep{Gao2022TevatronAE}.
For a fair comparison with our model, we re-implement the DPR baseline using the same set of hyperparameters.

We train all the models on 8 NVIDIA Telsa V100 GPUs with 32GB memory.
We initialize all the encoders with \verb|bert-base-uncased|.
The max sequence length is 16 for query and 128 for passage.
The number of positive and negative passages follows a ratio of 1:7 for each sample.
We set the batch size to 32.
We use both officially provided BM25 negatives and in-batch negatives to train the models.
We use Adam optimizer with the learning rate of $5\times 10^{-6}$, linear decay with 10\% warmup steps.

In the preliminary study without data augmentation, we train both models for 10 epochs.
To make full use of generated queries, we first pre-train the models for 10 epochs on the corpus with a batch size of 256 and only in-batch negatives, and then fine-tune the models for 20 epochs till convergence on the training set.
We haven't tuned other hyperparameters.
The pre-training stage takes about 15 hours and the fine-tuning stage takes about 8 hours.

During inference, we use \verb|IndexFlatIP| of the \verb|faiss| library \citep{8733051} to index the corpus and perform an exact search.

\subsection{Results}

Table~\ref{tab:msmarco} illustrates the evaluation results of our model and the baselines.

\begin{table*}[ht]
    \centering
    \begin{tabular}{c|c|cc|c|c}
    \hline
    \multirow{2}*{Model} & \multirow{2}*{PLM} & \multicolumn{2}{c|}{MS MARCO} & TREC DL 19 & TREC DL 20 \\
    \cline{3-6}
     & & MRR@10 & Recall@1k & nDCG@10 & nDCG@10 \\
    \hline\hline
    \multicolumn{6}{c}{\textbf{Sparse}} \\
    \hline
    BM25 & - & 18.4 & 85.3 & 50.6 & 48.0 \\
    DocT5Query & - & 27.7 & 94.7 & 64.8 & 61.9 \\
    DeepImpact & BERT$_{base}$ & 32.6 & 94.8 & 69.5 & 65.1 \\
    \hline\hline
    \multicolumn{6}{c}{\textbf{Dense}} \\
    \hline
    DPR & BERT$_{base}$ & 31.4 & 95.3 & 59.0 & 62.1 \\
    ANCE & RoBERTa$_{base}$ & 33.0 & 95.9 & 64.8 & - \\
    ME-BERT & BERT$_{large}$ & 33.4 & - & 68.7 & - \\
    DRPQ & BERT$_{base}$ & 34.5 & 96.4 & - & - \\
    ColBERT & BERT$_{base}$ & 36.0 & 96.8 & 69.4 & 67.6 \\
    \hline
    Ours & BERT$_{base}$ & 36.0 & 96.4 & 68.3 & 68.9 \\
    \hline
    \end{tabular}
    \caption{Evaluation results on MS MARCO passage retrieval benchmark and TREC DL track.
    DocT5Query and DeepImpact can be seen as the sparse counterparts of our model.
    Both ME-BERT and DRPQ learn multi-vector encodings for documents, and have used the hard negative mining technique proposed in ANCE.
    ColBERT learns term-level representations of both query and document for late interaction.
    Results not available are marked as `-'.
    }
    \label{tab:msmarco}
\end{table*}

We first compare our model against the DPR dual encoder baseline.
We can observe substantial improvements in terms of MRR@10 and nDCG@10 across all these datasets, which demonstrate the effectiveness of our approach.
The Recall@1k also exhibits a slight improvement.

Our approach is also competitive with other baselines.
On MS MARCO, it surpasses other baselines and is comparable to ColBERT, while being more efficient.
On TREC DL 19, the results are comparable to ME-BERT, which used a more powerful large-size model as backbone and the hard negative mining technique of ANCE.
On TREC DL 20, our model even outperforms the ColBERT model.

\subsection{Ablation Study}

We conduct ablation studies on our model design choice.

\subsubsection{Effect of Data Augmentation}

We used the generated queries as data augmentation for pre-training.
We ablate on the effect of pre-training in this section.
The results of different training stages on MS MARCO dev set are shown in Table~\ref{tab:pretrain}.

\begin{table}[ht]
    \centering
    \begin{tabular}{c|c|c|c}
    \hline
    MRR@10 & Pretrain & Finetune & Full \\
    \hline\hline
    DPR & 25.6 & 31.4 & 34.2 \\
    \hline
    Ours & 26.1 & 33.2 & 36.0 \\
    \hline
    \end{tabular}
    \caption{Ablation of different training stages on MS MARCO dev set. Pretrain: only use generated data for training; Finetune: only use data from training set for training; Full: Pretrain + Finetune.}
    \label{tab:pretrain}
\end{table}

We can see that using generated data for pre-training gives a MRR@10 score comparable to DocT5Query but lower than directly fine-tuning using data from the training set.
The top-$k$ sampling decoding strategy in query generation may introduce some noise, which explains why the pre-training underperforms directly fine-tuning with high-quality data.
However, the pre-training stage is still beneficial for the fine-tuning stage.

The results on TREC DL track are shown in Table~\ref{tab:trec}.
Our model still consistently outperforms the dual encoder baseline under different settings.
The improvements are more significant on this benchmark since the annotation is more complete.
Notably, our model without data augmentation is comparable to the DPR baseline with data augmentation on this benchmark.

\begin{table}[!ht]
    \centering
    \begin{tabular}{c|cc}
    \hline
    nDCG@10 & DL 19 & DL 20 \\
    \hline\hline
    \multicolumn{3}{c}{w/o Data Augmentation} \\
    \hline
    DPR & 59.0 & 62.1 \\
    Ours & 63.0 & 67.6 \\
    \hline\hline
    \multicolumn{3}{c}{w/ Data Augmentation} \\
    \hline
    DPR & 63.1 & 66.5 \\
    Ours & 68.3 & 68.9 \\
    \hline
    \end{tabular}
    \caption{Results on TREC DL track under different settings.}
    \label{tab:trec}
\end{table}

\subsubsection{Effect of Sharing Parameters}

Sharing the encoder parameters can reduce the number of model parameters to half.
We tie our encoder parameters in main experiments but also provide ablation of untied encoder parameters in Table~\ref{tab:untie} to study its effect.

\begin{table}[!ht]
    \centering
    \begin{tabular}{c|cc}
    \hline
    MRR@10 & tie & untie \\
    \hline\hline
    DPR & 31.4 & 31.7 \\
    Ours & 33.2 & 33.8 \\
    \hline
    \end{tabular}
    \caption{Results of tie / untie encoder parameters on MS MARCO dev set.}
    \label{tab:untie}
\end{table}
We can observe that using two sets of encoder parameters gives slightly better performance but not so significantly. 
Using separate encoders brings more improvements to our model, which is normal since the nature of two encoders in our model is more asymmetric than that in the vanilla dual encoder.

\section{Analysis}

Our experimental results in the previous section demonstrate that it is indeed beneficial to incorporate query interactions into the document representations.
The generated queries are crucial to the success of our model.
In this section, we analyse the influence of query quality and query diversity to the retrieval performance.

\subsection{On the Query Quality}

The number of queries is an important factor in our framework.
Too few queries have low diversity while too many queries will sacrifice efficiency.
Thus we provide an analysis here to study its effect.
We evaluate the query generation performance on the dev set of MS MARCO and reveal its relation with the retrieval performance.

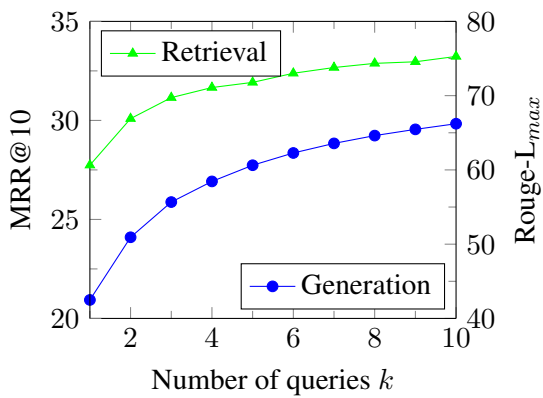
\begin{figure}[!ht]
    \centering
\pgfplotstableread{rouge-mrr.dat}{\table}
\begin{tikzpicture}
\begin{axis}[
    xmin = 1, xmax = 10,
    ymin = 40, ymax = 80,
    axis y line*=right,
    xlabel = {Number of queries $k$},
    ylabel = {Rouge-L$_{max}$},
    ylabel near ticks,
    minor tick num = 1,
    major grid style = {lightgray},
    minor grid style = {lightgray!25},
    width = 0.4\textwidth,
    legend cell align = {left},
    legend pos = south east,
]

\addplot[blue, mark = *] table [x = {k}, y = {rouge-l}] {\table};
\legend{
Generation
}
\end{axis}

\begin{axis}[
    xmin = 1, xmax = 10,
    ymin = 20, ymax = 35,
    hide x axis,
    axis y line*=left,
    ylabel = {MRR@10},
    ylabel near ticks,
    minor tick num = 1,
    major grid style = {lightgray},
    minor grid style = {lightgray!25},
    width = 0.4\textwidth,
    legend cell align = {left},
    legend pos = north west,
]
\addplot[green, mark = triangle*] table [x = {k}, y = {mrr}] {\table};
\legend{
Retrieval
}
\end{axis}
\end{tikzpicture}
    \caption{The evolution of ROUGE-L and MRR@10 on MS MARCO dev set when varying number of queries from 1 to 10.}
    \label{fig:effect-k}
\end{figure}

To measure the generation quality, we calculate the maximum ROUGE-L score between generated queries and the gold query on the dev set.
For retrieval performance, we report the MRR@10.
Figure~\ref{fig:effect-k} illustrates the evolution of the two metrics with different number of queries.\footnote{
Please refer to Table~\ref{tab:num-query} in Appendix~\ref{sec:appendix} for exact numbers.
}

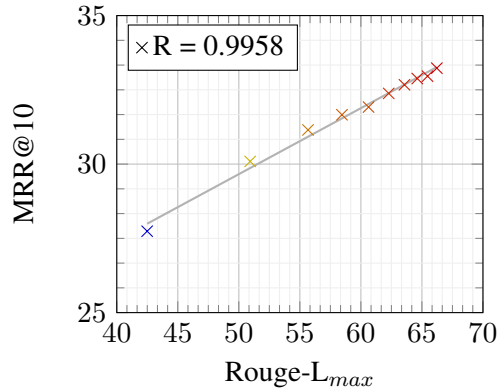
\begin{figure}[!ht]
\centering
\begin{tikzpicture}
\begin{axis}[
    xmin = 40, xmax = 70,
    ymin = 25, ymax = 35,
    xlabel = {Rouge-L$_{max}$},
    ylabel = {MRR@10},
    width = 0.4\textwidth,
    xtick distance = 5,
    ytick distance = 5,
    grid = both,
    minor tick num = 5,
    major grid style = {lightgray},
    minor grid style = {lightgray!25},
    legend cell align = {left},
    legend pos = north west,
]
\addplot[scatter, only marks, mark = x, mark size = 3pt] table[x={rouge-l}, y={mrr}] {rouge-mrr.dat};
\addplot[thick, black!30] table[
  x = {rouge-l},
  y = {create col/linear regression = {y = mrr}}
] {rouge-mrr.dat};
\legend{
R = 0.9958,
}
\end{axis}
\end{tikzpicture}
    \caption{Correlation between generation and retrieval performance on MS MARCO dev set.}
    \label{fig:correlation}
\end{figure}

We can see that as the number of queries grows, the retrieval performance becomes better because of the improved generation quality.
The correlation between the two metrics is shown in Figure~\ref{fig:correlation}.
The Pearson coefficient is 0.9958, indicating a strong positive correlation.
Keep increasing the number of queries will consistently improve the retrieval performance but more marginally.

\subsection{On the Query Diversity}
Intuitively, more diverse queries can potentially hit more types of queries.
We used top-$k$ sampling strategy to encourage the query generation diversity.
However, whether and to what extent the generated queries are diverse remains unclear.
To this end, we adopt the self-BLEU \citep{DBLP:conf/sigir/ZhuLZGZWY18} to measure the query generation diversity for a document.

\begin{figure*}[!ht]
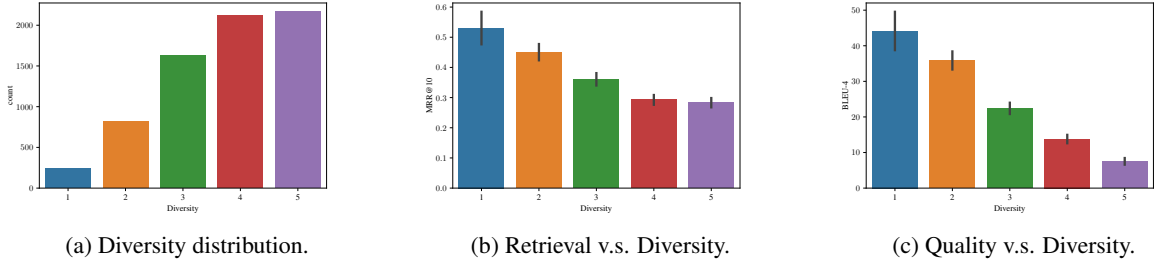

    \centering
\begin{subfigure}{0.32\textwidth}
    \scalebox{0.32}{\input{diversity-histogram.pgf}}
    \caption{Diversity distribution.}
    \label{fig:diversity-dist}
\end{subfigure}
\hfill
\begin{subfigure}{0.32\textwidth}
    \scalebox{0.32}{\input{mrr-diversity.pgf}}
    \caption{Retrieval v.s. Diversity.}
    \label{fig:mrr-diversity}
\end{subfigure}
\hfill
\begin{subfigure}{0.32\textwidth}
    \scalebox{0.32}{\input{diversity-quality.pgf}}
    \caption{Quality v.s. Diversity.}
    \label{fig:diversity-quality}
\end{subfigure}
\caption{Statistics of query diversity on MS MARCO dev set. We divided the diversity into 5 levels based on an average division of the self-BLEU-4 score.}
\label{fig:diversity}
\end{figure*}

We partition the documents of MS MARCO dev set to subsets of different query diversity according to their self-BLEU-4 score and measure the retrieval performance on these subsets.
The statistics are shown in Figure~\ref{fig:diversity}.
First, we observe that most of the documents have high query generation diversity thanks to the top-$k$ sampling strategy (see Figure~\ref{fig:diversity-dist}).
Second, the retrieval performance drops with higher diversity (see Figure~\ref{fig:mrr-diversity}).
One possible reason is that the QG model will generate more diverse queries when it doesn't know the right one.
As such, higher diversity indicates lower quality (see Figure~\ref{fig:diversity-quality}) and the retrieval performance drops with lower generation quality (see Figure~\ref{fig:correlation}).
It would be desirable to design a diversity metric that takes into account the generation quality.

\subsection{Case Study}

\newcommand{\tabincell}[2]{\begin{tabular}{@{}#1@{}}#2\end{tabular}}

\begin{table*}[ht]
    \centering
    \begin{tabularx}{\textwidth}{l|X}
    \hline
    Query & how old is canada \\
    \hline\hline
    Ours Rank 1 & Canada was finally established as a country in 1867. It is 148 years old as of July 1 2015. Canada has been a country for 147 years. The first attempt at colonization occurred in 1000 A.D. by the Norsemen. There was no further European exploration until 1497 A.D. when the Italian sailor John Cabot came along. It then started being inhabited by more Europeans. \\
    \hline
    Generated Queries & \tabincell{l}{
    when was canada established \\
    when was canada discovered \\
    what year was canada founded \\
    how long has canada been a country \\
    how old is canada
    } \\
    \hline\hline
    DPR Rank 1 & it depends where you live but in Canada you have to be at least 16 years old. \\
    \hline
    Generated Queries & \tabincell{l}{
    what is the legal age to be in canada \\
    how old do you have to be to live in canada \\
    how old do you have to be to enter canada as a citizen \\
    at what age can i go to canada to study in canada \\
    what is the minimum age to join the military
    } \\
    \hline
    \end{tabularx}
    \caption{Case Study on MS MARCO dev set.}
    \label{tab:case-study}
\end{table*}

We conduct a case study on the dev set to intuitively compare our model and the dual encoder baseline, as well as to illustrate the QG performance.

Table~\ref{tab:case-study} shows an example drawn from the MS MARCO dev set.
Our model can retrieve the correct passage by generating the right query.
DPR retrieves a hard negative passage where the content is corresponding to the query keywords but can not correctly answer the query.
By generating queries, our model can better distinguish the difference among document meanings.

\section{Discussion}

The ranking task is usually approached with a two-stage pipeline: retrieve-then-rerank.
The two stages usually use different architectures due to the effectiveness and efficiency trade-off.
Dual encoder is more efficient for retrieval, while cross encoder is more powerful for reranking.
How to take advantage of each other's strengths for mutual improvements is a hot topic of research.
We propose a new dual cross encoder architecture to benefit from both with a pre-interaction mechanism.

One limitation of our framework is that there exists a discrepancy between training and inference.
We used the gold query to train the model but do not have access to the gold query during inference.
Generating more queries would bridge this gap, but at the cost of efficiency.
We wish to close this gap with improved training strategy or improved query generation quality in the future.

\section{Conclusion}

We proposed a novel dense retrieval model to bridge the gap between dual encoder and cross encoder.
In our framework, the document representations are obtained by pre-interacting with a set of generated pseudo-queries through a cross encoder.
Our approach enables multi-view document representation with deep query interaction while maintaining the inference efficiency of the dual encoder approach.
We demonstrated its effectiveness compared to dual encoder baseline via experiments on various retrieval benchmarks.
In the future work, we would like to explore how to better incorporate generated queries for model training and how to improve the query generation quality for better retrieval performance.


\bibliography{anthology,custom}
\bibliographystyle{acl_natbib}

\appendix

\section{Appendix}
\label{sec:appendix}

\begin{table}[!ht]
    \centering
    \begin{tabular}{c|cc}
    \hline
    $k$ & ROUGE-L & MRR@10 \\
    \hline
    1 & 42.49 & 27.74 \\
    2 & 50.93 & 30.09 \\
    3 & 55.67 & 31.15 \\
    4 & 58.45 & 31.66 \\
    5 & 60.63 & 31.92 \\
    6 & 62.28 & 32.38 \\
    7 & 63.57 & 32.67 \\
    8 & 64.62 & 32.88 \\
    9 & 65.46 & 32.96 \\
    10 & 66.22 & 33.23 \\
    \hline
    \end{tabular}
    \caption{Results of generation and retrieval performance on MS MARCO dev set when varying number of queries (correspond to Figure~\ref{fig:effect-k} and Figure~\ref{fig:correlation}).}
    \label{tab:num-query}
\end{table}

\end{document}